\documentstyle[11pt,amssymb]{article}

\def\dalemb#1#2{{\vbox{\hrule height .#2pt
        \hbox{\vrule width.#2pt height#1pt \kern#1pt
                \vrule width.#2pt}
        \hrule height.#2pt}}}
\def\square{\mathord{\dalemb{6.8}{7}\hbox{\hskip1pt}}}

\def\0{{\sst{(0)}}}
\def\1{{\sst{(1)}}}
\def\2{{\sst{(2)}}}
\def\3{{\sst{(3)}}}
\def\4{{\sst{(4)}}}
\def\5{{\sst{(5)}}}
\def\6{{\sst{(6)}}}
\def\7{{\sst{(7)}}}
\def\8{{\sst{(8)}}}

\def\ep{\epsilon}
\def\td{\tilde}
\def\wtd{\widetilde}

\let\a=\alpha

\def\nn{\nonumber} \def\bd{\begin{document}} \def\ed{\end{document}}
\def\ds{\documentstyle} \let\fr=\frac \let\bl=\bigl \let\br=\bigr
\let\Br=\Bigr \let\Bl=\Bigl 
\let\bm=\bibitem
\let\na=\nabla
\let\pa=\partial \let\ov=\overline 
\newcommand{\be}{\begin{equation}} 
\newcommand{\ee}{\end{equation}} 
\def\ba{\begin{array}}
\def\ea{\end{array}}
\def\ft#1#2{{\textstyle{{\scriptstyle #1}\over {\scriptstyle #2}}}}
\def\fft#1#2{{#1 \over #2}}
\def\del{\partial}
\def\sst#1{{\scriptscriptstyle #1}}
\def\oneone{\rlap 1\mkern4mu{\rm l}}
\def\ie{{\it i.e.\ }}
\def\via{{\it via}}
\def\semi{{\ltimes}}
\def\str{{\rm str}}
\def\jm{{\rm j}}
\def\im{{\rm i}}
\def\bOmega{{{\bar\Omega}}}
\def\Qn{{{Q_{\sst{\rm N}}}}}
\def\tX{{{\wtd X}}}

\def\mapright#1{\smash{\mathop{-\!\!\!-\!\!\!-\!\!\!-\!\!\!-\!\!\!
             \longrightarrow}\limits^{#1}}}
\def\maprightt#1#2{\smash{\mathop{-\!\!\!-\!\!\!-\!\!\!-\!\!\!-\!\!\!
             \longrightarrow}\limits^{#1}_{#2}}}

\newcommand{\ho}[1]{$\, ^{#1}$}
\newcommand{\hoch}[1]{$\, ^{#1}$}
\newcommand{\bea}{\begin{eqnarray}} 
\newcommand{\eea}{\end{eqnarray}} 
\newcommand{\ra}{\rightarrow}
\newcommand{\lra}{\longrightarrow}
\newcommand{\Lra}{\Leftrightarrow}
\newcommand{\ap}{\alpha^\prime}
\newcommand{\bp}{\tilde \beta^\prime}
\newcommand{\tr}{{\rm tr} }
\newcommand{\Tr}{{\rm Tr} }

\newcommand{\NP}{Nucl. Phys. }
\newcommand{\tamphys}{\it Center for Theoretical Physics\\
Texas A\&M University, College Station, Texas 77843}
\newcommand{\ens}{\it Laboratoire de Physique Th\'eorique de l'\'Ecole
Normale Sup\'erieure\hoch{2,3}\\
24 Rue Lhomond - 75231 Paris CEDEX 05}
\newcommand{\upenn}{\it Department of Physics and Astronomy\\
University of Pennsylvania, Philadelphia, Pennsylvania 19104}

\newcommand{\auth}{M. Cveti\v{c}\hoch{\dagger1}, 
H. L\"u\hoch{\dagger1} and C.N. Pope\hoch{\ddagger2}}

\begin{document}
\begin{flushright}
\hfill{CTP TAMU-24/00}\\
\hfill{UPR-896-T}\\
\hfill{hep-th/0007209}\\
\hfill{July, 2000}\\
\end{flushright}


\begin{center}
{ \large {\bf Domain Walls with Localised Gravity and \\
Domain-Wall/QFT Correspondence }}

\vspace{15pt}
\auth

\vspace{15pt}

{\hoch{\dagger}\upenn}

\vspace{15pt}
{\hoch{\ddagger}\tamphys}

\vspace{40pt}

\underline{ABSTRACT}
\end{center}

     We review general domain-wall solutions supported by a
delta-function source, together with a single pure exponential scalar
potential in supergravity.  These scalar potentials arise from a
sphere reduction in M-theory or string theory.  There are several
examples of flat (BPS) domain walls that lead to a localisation of
gravity on the brane, and for these we obtain the form of the corrections to
Newtonian gravity. These solutions are lifted back on certain internal
spheres to $D=11$ and $D=10$ as M-branes and D-branes.  We find that
the domain walls that can trap gravity yield M-branes or D$p$-branes
that have a natural decoupling limit, {\it i.e.} $p\le 5$, with the
delta-function source providing an ultra-violet cut-off in a dual
quantum field theory.  This suggests that the localisation of gravity
can generally be realised within a Domain-wall/QFT correspondence,
with the delta-function domain-wall source providing a cut-off from
the space-time boundary for these domain-wall solutions.  We also
discuss the form of the one-loop corrections to the graviton
propagator from the boundary QFT that would reproduce the corrections
to the Newtonian gravity on the domain wall.

{\vfill\leftline{}\vfill
\footnoterule
{\footnotesize \hoch{1} Research supported in part by DOE grant 
DE-FG02-95ER40893 and NATO grant 976951. \vskip -12pt} \vskip 14pt
{\footnotesize  \hoch{2} Research supported in part by DOE 
grant DE-FG03-95ER40917.\vskip  -12pt}}

\pagebreak
\setcounter{page}{1}

\section{Introduction}

    The most intriguing feature of the Randall-Sundrum II scenario
\cite{rs2}
is the observation that a five-dimensional theory with a non-compact
fifth direction can nevertheless describe four- dimensional gravity on
a domain wall.  In the original formulation \cite{rs2}, the
configuration is a flat infinitely thin $Z_2$ symmetric domain wall
with with an asymptotically AdS space-time.\footnote{Smooth flat
domain-wall configurations of this type were first found in $D=4$
\cite{cgr} as BPS configurations interpolating between supersymmetric
isolated extrema in $N=1$ supergravity theory. Subsequent
generalisations led to the study of the global and local space-time
structures of non-supersymmetric thin-wall configurations (bent
domain-walls) \cite{cgs} as well as dilatonic domain walls
\cite{cvetic9402089,cs1} and were reviewed in \cite{cs2}.}  Such a
domain wall can be viewed as a solution of five-dimensional (AdS$_5$)
supergravity in the bulk on either side of the 3-brane (tending to the
AdS$_5$ horizon far from the brane), but it requires a delta-function
source on the 3-brane.  The origin of this singular source lies
outside five-dimensional supergravity, and within a field-theoretic
discussion, it is put by hand.  Smooth supergravity solutions that
exhibit the same feature of localising gravity on the brane remain
elusive, and indeed no-go theorems have been established that
demonstrate their absence, subject to certain sets of assumptions
\cite{bc1,zukov,kalshmak,kallin,bc2,giblam,ceresole}.  (For related
work see also \cite{hver,verlindetal,manu}.)

     In another development, it was suggested that there could be a
complementarity between the localisation of gravity in the
Randall-Sundrum scenario and the AdS/CFT correspondence
\cite{hver,witmaldunpub,gub1,gidkatran,noz,hawherrea}.  In the
Randall-Sundrum scenario the introduction of the singular domain-wall
(3-brane) source in the AdS space-time, before the boundary of AdS is
reached, can be viewed within the AdS/CFT correspondence as the
introduction of an ultra-violet cut-off in the boundary conformal
filed theory, thus introducing gravity on the domain wall
\cite{witmaldunpub}.  An important test of this complementarity is to
see whether the leading-order corrections to Newtonian gravity on the
3-brane in the Randall-Sundrum picture are in agreement with the
predictions from the corrections to the graviton propagator induced by
one-loop SCFT effects on the boundary
\cite{witmaldunpub,gub1,gidkatran}.  The matching between the two
approaches was recently demonstrated in \cite{duffliu}.

   It has been proposed that the conjectured AdS/CFT correspondence
\cite{malda,gkp,wit} can be extended to a more general class of
theories, where a bulk supergravity admitting a domain-wall solution
(describing a dilatonic vacuum) is dual to a quantum field theory on
the boundary of the solution at infinity \cite{towske,bergs1,clpuniv};
this is referred to as the Domain-wall/QFT correspondence.  It is
therefore natural to conjecture that under suitable circumstances
there could be a complementarity between this Domain-wall/QFT
correspondence and a localisation of gravity on the domain wall
itself.  In this paper we search for evidence that might support this
conjecture and show that within this complementarity framework the
localisation of gravity takes place precisely when a natural
decoupling limit within the Domain-wall/QFT correspondence can be
established.
 
    To do this, we begin by studying a large class of BPS domain-wall
solutions in supergravity theories, namely those that involve a single
scalar field that has a pure single-exponential potential. These bulk
Lagrangians are supplemented by a singular domain wall source.  In
section 2 we begin with a review of these solutions, including their
space-time structure and the matching across the domain-wall source.
By studying the equations governing linearised fluctuations of the
graviton, we then study the conditions for the localisation of
gravity on the wall.  We also obtain the leading-order corrections to
Newtonian gravity in those cases where the localisation occurs.

    In section 3 we study the origins of pure single-exponential
scalar potentials in supergravities, coming from reductions of
M-theory or string theory.  They can arise, for example, from
generalised Scherk-Schwarz reductions on Ricci-flat internal spaces,
or from reductions on internal Einstein spaces (such as spheres).  We
find that the Ricci-flat reductions never give rise to
gravity-trapping domain walls, but that under appropriate
circumstances the scalar potentials coming from spherical reductions
can give domain walls that do trap gravity. We classify the higher
dimensional origins of these solutions, and the circumstances under
which the gravity trapping solutions can appear in the
lower-dimensional theory.

    We find that the examples where gravity-trapping occurs can all be
lifted back to become the near-horizon regions of M-branes or
D$p$-branes with $p\le 5$.  These are precisely the branes for which a
natural gravity-decoupling limit exists, which is a {\it sine qua non}
for the possibility of establishing a Domain-wall/QFT correspondence.
For the D$p$-branes with $p\ge6$, on the other hand, it is known that
there is no natural decoupling limit \cite{sen,seiberg}, and
correspondingly, we find that in these cases there is no associated
domain wall that can trap gravity.  Thus we obtain strong evidence
that the localisation of gravity on the domain wall, and the existence
of a Domain-wall/QFT correspondence, go hand in hand.  We explore the
complementarity further in section 4, by considering the one-loop
corrections to the graviton propagator from the Yang-Mills fields on
the boundary.  Concluding remarks are contained in section 5.  In an
appendix, we obtain the exact analytic expression for the Green
function for the operator describing linearised gravity fluctuations,
for one specific example of a gravity-trapping domain wall.

\section{Supergravity domain walls and localisation of gravity}

\subsection{Supergravity domain-wall solutions}

     Our starting point is a $d$-dimensional action of the form
\be
S= \int_{\Sigma_d} \sqrt{-g}\,d^d x\, (R -\ft12 (\del\phi)^2 - 2\Lambda\,
e^{b\,\phi})  + \int_{\Sigma_{d-1}} d^{d-1}x\, {\cal L}_{\rm source}\,,
\label{1exp}
\ee
where the bulk contribution is viewed as arising from a supergravity
theory whose higher dimensional origin, as an effective theory of
sphere reduced M-theory or string theory, we shall discuss in the
subsequent section. To this bulk action we added by hand a
delta-function source for the domain wall; this delta function will
provide a cut-off for the boundary of the bulk solution. For the case
of the AdS bulk solution, this source provides a cut-off from the AdS
boundary (in horospherical coordinates) and thus within the AdS/CFT
correspondence provides the ultra-violet cut-off for the dual CFT
description of the brane-world scenario
\cite{hver,witmaldunpub,gub1,gidkatran,noz,hawherrea,duffliu}.

     Single-exponential scalar potentials in fact occur rather
frequently in such circumstances, as we shall discuss in detail in the
next section.\footnote{One could refer to such bulk Lagrangians as
``vacuum'' Lagrangians where one has not excited any other massless
scalar fields; see, for example, \cite{clpuniv}. Situations involving
additional fields can also be considered.} It is useful to
parameterise the constant $b$ that determines the strength of the
coupling in terms of a constant $\Delta$ as follows:
\be
b^2= \Delta + \fft{2(d-1)}{d-2}\,.\label{bdelta}
\ee
The advantage of this parameterisation is that the value of $\Delta$ is
preserved under toroidal dimensional reduction \cite{stainless}.  Note
that a pure cosmological term, corresponding to $b=0$, has
\be
\Delta = \Delta_{\rm AdS} \equiv -2 -\fft{2}{d-2} <-2\,.\label{deltaads}
\ee
The reality of $b$ requires that $\Delta$ satisfy
\be
\Delta \ge \Delta_{\rm AdS}\,.\label{realb}
\ee

   We shall review domain-wall solutions \footnote{These solutions
were first considered  \cite{cvetic9402089} as BPS solutions in
$d=4$.  Generalisations to $D$ dimensions in the context of localisation 
of gravity were studied in \cite{youm1,youm2}.} in the conformally-flat
frame
\be
ds^2 = e^{2A}\, (\eta_{\mu\nu}\, dx^\mu\, dx^\nu + dz^2)\,,
\label{conf}
\ee
where $A$ and the bulk scalar field $\phi$ is a function only of $z$. 

   In principle these solutions can be obtained as solutions to the
first order (BPS) differential equations (c.f.
\cite{cvetic9402089} for the $d=4$ case), with the Israel matching
conditions \cite{israel} across the singular domain wall source
relating the tension (${\cal T}= \int_{\Sigma_{d-1}} d^{d-1}x\, {\cal
L}_{\rm source}$) of the wall to the bulk cosmological constant
parameters. For the sake of simplicity we choose to have the {\it
same} bulk Lagrangian (\ref{1exp}) on either side ($z>0$ and $z<0$) of
the wall and thus the solution is $Z_2$ symmetric.\footnote{ The
choice of a different bulk Lagrangian on either side ($z>0$ or $z<0$)
of the wall, for example with a different choice of $\Lambda$ and/or
$b$ on either side of the wall, would in turn allow one to construct
$Z_2$ asymmetric walls.}  In addition we shall focus on the
positive tension walls (${\cal T}>0$) and only briefly discuss the
negative tension branes. (The latter turn out not to be of interest
for localisation of gravity.)

 For $\Delta\ne -2$ the equations of motion following from
(\ref{1exp}) admit domain-wall solutions, given by
\bea
ds^2 &=& H^{\fft{4}{(d-2)(\Delta+2)}}\, (\eta_{\mu\nu}\, dx^\mu\,
dx^\nu + dz^2)\,,\nn\\
e^\phi &=& H^{-\fft{2b}{\Delta+2}}\,,\qquad H = 1+ k\, |z|\,,\label{gensol}
\eea
where
\be
k^2 = \fft{(\Delta+2)^2\, \Lambda}{\Delta}\,.\label{ksquared}
\ee
In order to have a real solution, it is necessary that $\Lambda$ and
$\Delta$ have the same sign.  As we shall discuss in the next section,
this is indeed always the case in supergravity theories.  The constant
$k$ can then be chosen to be either the positive or the negative root
of (\ref{ksquared}). In the former case the $z$ coordinate runs from
$-\infty$ to $+\infty$ while in the latter the range of the transverse
coordinate is finite. The choice of the sign of $k$ has to be
correlated with the matching condition across the singular domain wall
source.

    If $\Delta=-2$, the solution is instead given by
\bea
ds^2 &=& e^{-\fft{2k}{d-2} \, |z|}\, (\eta_{\mu\nu}\, dx^\mu\, dx^\nu +
dz^2)\,,\nn\\
\phi &=& \fft{\sqrt2\, k}{ \sqrt{d-2}}\, |z|\,,\label{d2sol}
\eea
where $k$ is now given by
\be
k^2 = -2\Lambda\, (d-2)\,,
\ee
which is real for negative $\Lambda$.

     The matching conditions across the singular domain wall source
imply that that the energy density (tension) of the wall is related to
the values of the cosmological constant parameters on either side of
the wall. (For a detailed discussion pertinent to our situation, see,
e.g., Refs. \cite{cs1,cs2}.) For the domain wall source associated
with a typical solitonic kink the matching conditions is of the form
\be
\sigma = {\cal T}=   2 (A'_{z=0^-}-A'_{z=0^+})\,,
\ee
where the prime denotes a derivative with respect to $z$.  This
leads to
\bea
\Delta\ne -2:&&  {\cal T}=
-8\, {\rm sign}[k\, (\Delta+2)]\sqrt{\Lambda\over \Delta}\,,\nn\\
\Delta=-2: && {\cal T} = \fft{8k}{d-2}\,.  
\label{tension}\eea
Thus positive-tension domain-wall solutions exist for $\Delta \le -2$
with $k>0$ and for $\Delta>-2$ with $k<0$.  Conversely,
negative-tension domain walls arise for $\Delta \le -2$ with $k<0$ and
for $\Delta>-2$ with $k>0$.  From now on, whenever we discuss domain
walls with $\Delta\le -2$, the lower bound (\ref{realb}) is to be
understood.

    The Riemann curvature of the metric (\ref{conf}) is of the form
$A''\, e^{-2A}$ or $(A')^2\, e^{-2A}$.  For the domain-wall solutions
given in (\ref{gensol}), these functions are of the form $H^{-2 -
\fft{4}{(d-2)(\Delta +2)}}$.  It follows that for the positive
positive tension solutions with $\{ \Delta_{\rm AdS} < \Delta \le
-2,\, k>0\}$ and $\{ \Delta>-2, \, k<0 \}$, there are curvature
singularities at $z=\pm \infty$ and $z=\pm {1\over {|k|}}$,
respectively\footnote{For special values of $\Delta$, such as
$\Delta=+1$ in $d=4$, the Ricci scalar $R$ may be finite but then
$R_{\mu\,\nu}R^{\mu\,\nu}$ is infinite \cite{cvetic9402089}.}.  Thus
the positive-tension solutions have a null singularity, coinciding
with the horizon, for $\Delta \le -2$, and have naked singularities
for $\Delta>-2$. (Of course the AdS example with $b=0$ is non-singular
at $z=\pm\infty$, corresponding to the AdS Cauchy horizon.)

  The negative-tension solutions with $\{ \Delta_{\rm AdS} < \Delta
\le -2,\, k<0\}$ have naked singularities at $z=\pm{1\over {|k|}}$,
while those with $\{ \Delta>-2, k>0\}$ are geodesically complete with
$z=\pm\infty$ corresponding to the boundary of space-time.

\subsection{Localisation of gravity on domain walls}

   Our focus is to identify within the framework a class of $Z_2$
symmetric domain-wall solutions without naked singularities that can
localise gravity.  A necessary condition for localised gravity on the
brane at $z=0$ is that the conformal scale factor $e^{2A}$ in
(\ref{conf}) should vanish at large $|z|$ and that the delta-function
source have a positive tension, thus providing a trapping volcano-like
effective potential at $z=0$.  For the solutions described above, this
happens only for the $\Delta\le -2$ and $k>0$ cases, \ie
$Z_2$-symmetric domain walls with positive tension, whose space-time
geometry has at most null singularities at $z=\pm\infty$. To see the
localisation of gravity in detail, one can examine the equation for
small gravitational fluctuations on the brane.  The fluctuations of
the $d$-dimensional graviton, in the conformal frame, satisfy the
equation of a minimally-coupled scalar field in the gravitational
background, namely $\del_M\,(\sqrt{-g}\, g^{MN}\, \del_N\, \Phi) =
0$. (See, {\it e.g.}, \cite{BS,csaki}.)  Following \cite{rs2}, we
consider the Ansatz
\be
\Phi = \phi(z)\, e^{\im\, p\cdot x} = e^{-\fft12 (d-2)\, A}\,
\psi(z)\, e^{\im\, p\cdot x}\,,
\ee
where the $A$-dependent rescaling function is chosen so that $\psi$
satisfies a Schr\"odinger-type equation,
\be
-\ft12 \psi'' + U\, \psi = -\ft12 p^2\, \psi\,,\label{schrod}
\ee
where the Schr\"odinger potential is given by
\bea
\Delta\ne -2:&& U = -\fft{(\Delta+1)\, k^2}{2(\Delta+2)^2\, H(z)^2}\, +
\fft{k}{\Delta+2}\, \delta(z)\,,\nn\\
\Delta=-2:&& U = \ft18 k^2\, - \ft12\, k\, \delta(z)\,.\label{upot}
\eea
The potential for $\Delta\ne -2$ was obtained in \cite{youm1,youm2}.

    It is evident from these expressions for $U$ that there will be a
zero-mass bound state if $\Delta \le -2$, since then the delta
function has a negative coefficient, and the ``bulk'' term is
non-negative for all $z$.  (Had we chosen the constant $k$ to be
negative, the solutions would have had naked singularities, and hence
will not be considered.)  In fact the potential is volcanic for
$\Delta<-2$, whilst for $\Delta=-2$ it is a raised constant potential,
again with a negative delta function.  The massless wave-function is
given by
\bea
\Delta <-2: && \psi = e^{\fft12 (d-2)\, A} =
H^{\fft{1}{\Delta+2}}\,,\nn\\
\Delta=-2: && \psi = e^{\fft12 (d-2)\, A}=e^{-\fft12 k\, |z|}\,.
\label{wavefunctions}
\eea
The trapping of gravity requires \cite{csaki} that the wave-function
be normalisable, \ie $\int|\psi|^2\, dz < \infty $, in order to obtain
a finite leading-order contribution to the gravitational field on the
brane.  This condition is satisfied, for the domain walls without
naked singularities that we are considering in this paper, if $-4 <
\Delta \le -2$.  In view of the bound (\ref{realb}) resulting from the
requirement that the constant $b$ be real, we therefore have the
criterion, for $d\ge4$, that
\be
-2-\fft{-2}{d-2}\equiv \Delta_{\rm AdS} \le  
        \Delta \le -2\,.\label{deltarange}  
\ee
The pure AdS case $b=0$ leads to a trapping of gravity \cite{rs2},
and indeed we see from (\ref{deltaads}) that the value of $\Delta$ in
this case lies strictly within the bound (\ref{deltarange}), for $d\ge
4$.  When $d=3$, we have $\Delta_{\rm{AdS}_3}=-4$, and the norm of the
wave-function is logarithmically divergent.  This may be attributed to
the degeneracy of two-dimensional pure gravity.  In $d \ge 4$ the
bound (\ref{realb}) on $\Delta$ that is needed for reality of the
exponential potential ensures that the $\Delta=-4$ limit in
(\ref{deltarange}) is never attained.
   
   It is worth noting that for $\Delta <-2$, including the AdS case,
there is no mass gap in the spectrum (although, of course, the
wavefunctions with small mass are delocalised away from the brane).
On the other hand, when $\Delta=-2$ there is a distinct mass gap, with
the continuum wavefunctions having $m^2=-p^2 \ge \ft14 k^2$.
Generally, for $\Delta$ approaching $-2$ from below, the effect of the
localisation of gravity becomes more pronounced. 

    For $\Delta$ lying outside the range (\ref{deltarange}), there
will be no trapping of gravity on the brane (for the examples with no
naked singularities that we are considering here).   In such cases one would
have to resort to the more traditional Kaluza-Klein approach of
compactifying the $z$ coordinate.  This can be done by introducing a
second parallel brane \cite{hw,losw,rs1}.

    In the next section we shall investigate the various exponential
potentials of the form (\ref{1exp}) that can arise in supergravities,
and in particular, their associated values of $\Delta$.

\subsection{Corrections to Newtonian gravity}

    It is of interest to see how the leading-order Newtonian
gravitational potential in the brane is modified for the various
five-dimensional domain walls that we have found.  In section 2, the
massless bound-states (\ref{wavefunctions}) were found.  Here, we
shall obtain the wavefunctions describing massive gravity fluctuations
also, and use these results in order to estimate the corrections to
the leading-order Newtonian result.

   The Schr\"odinger equation (\ref{schrod}) can be solved exactly for
the exponential scalar potentials that we are considering here.  After
imposing the boundary conditions at $z=0$, we find that the $Z_2$-symmetric
massive wavefunctions are given by
\bea
\Delta < -2: && \psi_m = c_m\, H^{1/2}\, \Big[ Y_{\nu-1}\Big(\fft{m}{k}\Big)\,
J_\nu\Big( \fft{m}{k}\, H\Big) - J_{\nu-1}\Big(\fft{m}{k}\Big)\,
Y_\nu\Big( \fft{m}{k}\, H\Big)\Big]\,,\nn\\
&& c_m = \sqrt{\fft{m}{\pi}}\, \Big[ Y_{\nu-1}\Big(\fft{m}{k}\Big)^2 +
 J_{\nu-1}\Big(\fft{m}{k}\Big)^2\Big]^{-1/2}\,,\\
&& \nu = \fft{\Delta}{2(\Delta+2)}\,,\nn\\
&&\nn\\
\Delta=-2: && \psi_m = c_m\, \Big( k\, \sin q|z| - 2q\, \cos q|z|
\Big)\,,\nn\\
&&c_m = \fft{1}{2m \sqrt{\pi\, q}}\,,\label{delta2wave}\\
&&q = \sqrt{m^2-\ft14 k^2}\,.\nn
\eea
Note that in the latter case, $\Delta=-2$, there is a mass gap and so
we must have $m^2\ge \ft14 k^2$ for the massive wavefunctions.

     The corrections to the Newtonian gravitational potential between 
masses $M_1$ and $M_2$ can be estimated as follows \cite{BS}:
\be
U(r) \sim \fft{G_4\, M_1\, M_2}{r} + \fft{G_5\, M_1\, M_2}{r}\, 
\int_{{m_0^2}}^\infty d(m^2)\, \psi_m(0)^2\, e^{-m\, r}\,,\label{correction}
\ee
where $m_0$ is the lowest mass for the non-bound states.  (This will
be taken to be zero, except in the case where there is a mass gap.)
The four-dimensional and five-dimensional Newton constants are related
by $G_4 = k\, G_5$.  For $\Delta < -2$ we therefore find
\be
U(r) \sim  \fft{G_4\, M_1\, M_2}{r} \Big(1 + \fft{c}{(kr)^{2\nu-2}} +
\cdots\Big)\,,
\ee
where $c$ is some constant of order 1.  The two cases that are
relevant to our discussion in this paper are $\Delta=-\ft83$, giving
$\nu=2$, and $\Delta=-\ft{12}{5}$, giving $\nu=3$.  The former is the
standard AdS Randall-Sundrum II scenario, with the well-known $1/r^3$
correction to the Newtonian potential; the latter gives instead a
$1/r^5$ correction at leading order.

   For $\Delta=-2$, we find
\be
U(r) \sim  \fft{G_4\, M_1\, M_2}{r}  
\Big(1 + \fft{2e^{-\ft12 k\, r}}{\pi\, k}\,   \int_0^\infty dy\, 
\fft{(y\, (y+k))^{1/2}}{y+\ft12 k}\,
e^{-y\,r} \Big)\,,
\ee
from which we see that the leading-order corrections are of the form
\footnote{We thank Konstadinos Sfetsos for pointing out an error in
the normalisation factor in (\ref{delta2wave}) in an earlier
version of this paper, which has resolved a previous
discrepancy with the Green function result obtained in the Appendix.}
\be
U(r) \sim  \fft{G_4\, M_1\, M_2}{r}  
\Big[1 + \fft{2e^{-\ft12 k\, r}}{\sqrt\pi}\, 
 \Big( \fft{1}{(k\, r)^{3/2}} -\fft{9}{4(k\,
r)^{5/2}} +\fft{345}{32(k\, r)^{7/2}}  + \cdots\Big)\Big]
\,.\label{del2res}
\ee
This is a Yukawa-like modification to Newtonian gravity.  The
essential $e^{-k\, r/2}$ factor reflects that we are dealing with a
situation where there is a mass gap $\ft12 k$ separating the
zero-energy bound state and the massive continuum.  The formula
(\ref{correction}) is consistent with the exact Green function we
obtain in Appendix.  The determination of the precise constant
coefficients involves subtleties concerning the imposition of gauge
conditions on the metric perturbations \cite{gartan,gidkatran,colhol}.
(The corrections to Newtonian gravity in a different domain wall whose
spectrum has a mass gap, associated with a D3-brane distributed over a
disc, was discussed in \cite{BS}.)

\section{Higher-dimensional origin of the scalar potentials}

\subsection{Exponential scalar potentials from Scherk-Schwarz reductions}

   The largest dimension in which a supergravity has a scalar
potential is $D=10$, in the massive type IIA theory.  This has
$\Delta=4$.  One way to obtain a scalar potential in a lower dimension
is by performing a Scherk-Schwarz reduction of a supergravity without
any scalar potential, in which an $n$-form field strength is taken to
be proportional to some harmonic $n$-form on a Ricci-flat internal
manifold.  The first example of this kind was the reduction of type
IIB supergravity on $S^1$, where the ``1-form field strength'' $d\chi$
was taken to be $m\, dy$, where $y$ is the tenth coordinate.  In fact
it was shown \cite{bergs2} that this reduction of type IIB is T-dual to
the massive type IIA theory.  Scherk-Schwarz reductions on tori were
extensively studied in \cite{clpst}, and a complete classification of
their exponential scalar potentials was given in \cite{classp}.  The
basic value of $\Delta$ that arises for any individual exponential
term is always $\Delta=4$.  If $N$ exponentials are combined,
eliminating some scalar fields using the equations of motion, one is
left with an exponential with $\Delta=4/N$, where $2\le N\le 8$ (the
possible values of $N$ depend on the dimension $D$
\cite{classp}).\footnote{In fact all known exponential scalar
couplings in ungauged supergravities have $\Delta=4/N$.
Interestingly, this provides strong supporting evidence for the belief
that $D=11$ supergravity cannot have a cosmological term, since from
(\ref{deltaads}) this would have $\Delta=-20/9$.  Since this value
would be preserved under toroidal dimensional reduction
\cite{stainless}, it would also imply the existence of
lower-dimensional supergravities with this peculiar value of
$\Delta$.}

    In some cases the compactifying tori can be replaced by certain
other Ricci-flat manifolds, such as K3 or a Calabi-Yau or Joyce
manifold.  This was discussed extensively in \cite{llp}.  One of the
examples is the Scherk-Schwarz reduction of M-theory on a
6-dimensional Calabi-Yau manifold $Y$, with the 4-form field strength
residing in the 4'th cohomology class of $Y$.  The resulting scalar
potential, with $\Delta=4/3$, was used to construct a domain-wall
solution in $d=5$ \cite{losw} that attempts to provide a field-theoretic
realisation of the Ho\v rava-Witten \cite{hw} construction.  
In a limit where the Calabi-Yau manifold becomes an orbifolded
6-torus, the domain wall solution can be viewed as an intersection of
three 5-branes \cite{llp}.

   In all the examples of Scherk-Schwarz reduction on Ricci-flat
internal manifolds, the value of $\Delta$ is positive and in addition
$\Lambda>0$, (unless the higher-dimensional theory already has a
scalar potential with negative $\Delta$.)  As we discussed in the
previous section, a single domain wall supported by such an
exponential potential cannot trap gravity.  It is then necessary to resort
to the traditional Kaluza-Klein mechanism where the extra dimension is
compact.  One way to compactify the extra dimension is to take the
coordinate $z$ to lie on the interval $S^1/Z_2$ \cite{losw}, with a
domain wall at each endpoint.

\subsection{Exponential scalar potentials from sphere reductions}

    Here we show that values of $\Delta$ that are less than $-2$ can
commonly arise from sphere reductions of M-theory or string theory.
Well-known examples are the supergravities with pure cosmological
constants ($b=0$) that arise from the $S^4$ and $S^7$ reductions of
M-theory, and the $S^5$ reduction of type IIB supergravity.  In this
section we shall consider a general Lagrangian in $D$ dimensions given
by
\be
\hat {\cal L} = \hat e\, \hat R - \ft12\, \hat e\,(\del\phi_1)^2 - 
\fft1{2\, n!}\, \hat e\, e^{a\, \phi_1}\, \hat F_{\sst{(n)}}^2\,,
\label{genlag}
\ee
where in supergravity theories the constant $a$ is parameterised by
\be
a^2 = \fft{4}{N} - \fft{2(n-1)(D-n-1)}{D-2}\,,\label{aval}
\ee
and $N$ is an integer. The case $N=1$ can arise for all $n$-forms in
supergravity.  In particular, in $D=10$ or $D=11$ all the field
strengths have $N=1$ \cite{dkl}, and in fact all the individual field
strengths in all maximal supergravities have $N=1$.  The case $N=2$ 
can arise for 2-forms in $D\le9$, and 3-forms in $D\le6$, in
non-maximal supergravities.  $N=3$ can arise for 2-forms in $D\le5$,
and $N=4$ for 2-forms in $D\le4$.  

     We now perform the following consistent reduction on $S^n$, using
the Ansatz\footnote{Note that here we are always giving a magnetic
charge to the $n$-form field strength.  The case where the charge is
instead electric can be handled within this framework by dualising the
$n$-form to a $(D-n)$-form.  The case of the self-dual 5-form in type
IIB supergravity was discussed in \cite{bdlps}, and the final result
is of the same form as (\ref{spherelag}).  Also, in this section we
consider $n\ge 2$, since the $n=1$ case was the topic of section 3.1.}
\bea
d\hat s_D^2 &=& e^{-2\a\, \phi_2}\, ds_d^2 + g^{-2}\, 
e^{\fft{2(d-2)}{n}\, \a\, \phi_2}\, d\Omega_n^2\,,\nn\\
\hat F_{\sst{(n)}} &=& m\, g^{-n}\, \Omega_n\,,\label{spherered}
\eea
where $d\Omega_n^2$ is the metric on the unit $n$-sphere, 
$\Omega_n$ is its volume form, and 
\be
\a = - \sqrt{\fft{n}{2(d-2)(D-2)}}\,.\label{alval}
\ee
This gives the following Lagrangian in $d=D-n$ dimensions \cite{bdlps}: 
\be
e^{-1}\, {\cal L} = R - \ft12(\del\phi_1)^2 -\ft12 (\del\phi_2)^2 -
\ft12 m^2\, e^{a\, \phi_1 -2(d-1)\, \a\, \phi_2}  +n(n-1)\, g^2\,
e^{-\fft{2\,\, (D-2)}{n}\, \phi_2}\,.\label{spherelag}
\ee

    Calculating the values of $\Delta$ for the two exponential terms,
we find
\be
\Delta_m=\fft{4}{N} \,,\qquad  \Delta_g = -2 + \fft2{n}\label{2del}
\ee
respectively.  Note that, comparing with (\ref{1exp}), the sign of
$\Lambda$ is indeed the same as that of $\Delta$ for each term, so
that $\Delta\, \Lambda$ is non-negative for each term.  One can use
either of the exponential terms by itself to construct a domain wall
of the kind discussed in section 2, since one can turn off either of
the parameters $m$ and $g$, and then, if necessary, rotate the
dilatons $\phi_1$ and $\phi_2$ to give a single dilatonic scalar in
the remaining exponential, with the orthogonal (free) dilaton set to
zero. As we discussed in section 2 there can be no localised gravity
in either case, since the inequality (\ref{deltarange}) is not
satisfied.   

   If instead $m$ and $g$ are both taken to be non-vanishing, we can
still eliminate one of the two dilatonic scalars.  The two
exponentials then coalesce into one, with a different value of
$\Delta$.  To see this, we make an orthonormal transformation to new
scalars $(\phi,\varphi)$, defined by
\be
\phi_1 = -\phi\, \cos\beta - \varphi\, \sin\beta\,,\qquad 
\phi_2 = \phi\, \sin\beta - \varphi\, \cos\beta\,,
\ee
with
\be
\tan\beta = -\sqrt{\fft{a^2\, n\, (D-2)}{2(d-2)(n-1)}}\,.
\ee
The Lagrangian (\ref{spherelag}) now becomes
\be
e^{-1}\, {\cal L} = R - \ft12(\del\phi)^2 -\ft12 (\del\varphi)^2 -
   V\,,
\ee
where
\be
V\equiv e^{b\, \phi}\, \Big(
\ft12 m^2\, e^{c_1\, \varphi}  - n(n-1)\, g^2\,
e^{c_2\, \varphi}\Big) \,,\label{spherepot}
\ee
and
\bea
&&c_1 =\sqrt{\fft{8n}{N\, [2n -(n-1)\, N]}} \,,\qquad
c_2 = (n-1)\, \sqrt{\fft{2N}{n\, [2n-(n-1)\, N]}}
\,,\nn\\
&&b^2 = -\fft{4(n-1)}{2n-(n-1)\, N} + \fft{2(d-1)}{d-2}
\,.\label{bc1c2}
\eea
For $N=1$ and $N=2$, the constants $c_1$ and $c_2$ are real for all
$n$.  If $N\ge 3$, the degree $n$ of the $n$-form must be less than
$N/(N-2)$.

    In this section we shall consider just a 1-scalar solution of the 
form discussed in section 2.  To do this, we first solve the $\varphi$
equation by taking $\varphi=0$, which therefore implies $c_1\, m^2 =
n(n-1)\, c_2 \, g^2$.  We are then left with the potential
\be
V = \fft{2(n-1)^2\, g^2}{\Delta}\, e^{b\, \phi}\,,\label{v1}
\ee
where $\Delta$ characterises the strength of the dilaton coupling as
in (\ref{bdelta}), and, from (\ref{bc1c2}), is given by
\be
\Delta = -\fft{4(n-1)}{2n-(n-1)\, N} \,.\label{combdel}
\ee
This expression was also given in \cite{youm1}, where its implications
for higher-dimensional origin of gravity trapping domain walls were
discussed.  

        Comparing with (\ref{1exp}), we see that the quantity
$\Delta\, \Lambda$ is indeed always positive, as we stated in the
previous section.

    For the relevant values of $N$, we therefore find that $\Delta$ is
as follows:
\bea
N=1\,,\quad D\le 11:&& \Delta = -4 + \fft{8}{n+1}\,,\nn\\
N=2\,,\quad D\le 9 :&& \Delta = 2-2n\,,\nn\\
N=3\,,\quad D\le 5: && \Delta = 4 + \fft{8}{n-3}\,,\nn\\
N=4\,, \quad D=4: && \Delta = 2 + \fft{2}{n-2}\,.
\eea
In order to satisfy the inequality (\ref{deltarange}) that ensures the
trapping of gravity on the brane, we can therefore have $N=1$ with
$n\ge 3$, or $N=2$ with $n=2$.  

       Note that the disallowed value of $(N,n)=(1,2)$ corresponds to
a reduction in which a Kaluza-Klein 2-form field strength is taken to
be proportional to the volume form of an internal $S^2$.  The
higher-dimensional theory could therefore itself be viewed as an $S^1$
reduction of pure gravity.  Thus the above analysis shows that it is
not possible to construct a gravity-trapping domain wall purely within
an Einstein gravity theory (with no cosmological constant).

    For a concrete example that is not without physical interest, we
can enumerate the various ways of obtaining five-dimensional theories
that are capable of trapping gravity on a 3-brane.  They correspond to
reducing an appropriate ordinary massless supergravity in $D=10$, 9, 8
or 7 on $S^5$, $S^4$, $S^3$ or $S^2$, respectively.  Thus they can all
be viewed as coming, for example, from reductions of type IIB
supergravity, as follows:

\bigskip\bigskip
\centerline{
\begin{tabular}{|c|c|c|c|c|}\hline
   & $S^5$ & $S^1\times S^4$ & $T^2\times 
S^3$ & $T^3\times S^2$ \\ \hline\hline
$N$ & 1 & 1 & 1 & 2  \\ \hline
$\Delta$ & $-\ft83$  & $-\ft{12}{5}$ & $-2$ & $-2$  \\ \hline
\end{tabular}}
\bigskip

\noindent{\bf Table 1:}  The sphere reductions that give trapped
gravity on a 3-brane in $D=5$, lifted to type IIB.

\bigskip\bigskip

In the next section, we discuss the higher-dimensional interpretations
of these solutions, both in the type IIB and the type IIA and M-theory
pictures.

\subsection{Lifting of the domain walls to higher dimensions}

    The various domain-wall solutions of the previous section,
supported by the potential $V$ given in (\ref{v1}) can be lifted back
on the $n$-sphere to $D$ dimensions where they become the near-horizon
limits of $(D-n-2)$-branes.  (See also \cite{youm1}.) To see this,
consider the standard isotropic $(D-n-2)$-brane in $D$ dimensions with
metric
\bea
ds_D^2 &=& \hat H^{-\fft{(n-1)\, N}{D-2}}\, \eta_{\mu\nu}\, dx^\mu\,
dx^\nu + \hat H^{\fft{(d-1)\, N}{D-2}}\, (dr^2 + r^2\,
d\Omega_n^2)\,,\nn\\
e^{-\ft{2}{N\, a}\,\phi} &=& \hat H\,\qquad
\hat H = 1 + \fft{Q}{r^{n-1}}\,.\label{genbranes}
\eea
If we drop the ``1'' in the harmonic function $\hat H$, and make the
coordinate redefinition 
\be
r= \Big(1 + k\, |z|\Big)^{-\fft{2}{(n-1)\, N -2}}\,,
\ee
then after performing the reduction of the $D$-dimensional
$(D-n-2)$-brane metric on the $n$-sphere, using the reduction Ansatz
given in (\ref{spherered}), and making appropriate constant
rescalings, we obtain precisely the $d$-dimensional domain-wall
solution (\ref{gensol}).  Note that there is an exceptional case when
$(n-1)\, N=2$, corresponding to $\Delta=-2$, for which we shall have
instead
\be
r= e^{-\ft{k}{n-1}\, |z|}\,,
\ee
reproducing (\ref{d2sol}).

    For the cases of principal interest to us, with $\Delta\le -2$,
the regions where $z \longrightarrow \pm\infty$ correspond to
$r\longrightarrow 0$.  Thus in these cases the horizons of the
domain-wall solution in the lower dimension $d$ (\ie $z\longrightarrow
\pm \infty$, far from the brane at $z=0$) map into the horizon of the
brane in the higher dimension $D=d+n$.  By contrast, in the cases with
$\Delta> -2$ the relation between $r$ and $z$ is of the form $r\sim
(1+k\, |z|)^c$ where $c$ is a positive constant, and there is no
region in the lower-dimensional solution that maps to the horizon of
the higher-dimensional brane.

     As an example, the domain-wall solutions listed in Table 1 have
oxidation endpoints as follows:

\bigskip\bigskip
\centerline{
\begin{tabular}{|c|c|c|c|c|}\hline
$\Delta$ & $-\ft83$  & $-\ft{12}{5}$ & $-2$ & $-2$  \\ \hline
Oxidation& D3 & M5 (or D4)  & D5 & M5$\perp$M5 \\
Endpoint & on $S^5$ & on $T^2\times S^4$ (or $S^1\times S^4$) &
          on $T^2\times S^3$ & on $T^4\times S^2$ or K3$\times S^2$\\ \hline
\end{tabular}}
\bigskip

\noindent{\bf Table 2:} Ten or eleven-dimensional origins of the
$\Delta\le -2$ five-dimensional domain walls.  
\bigskip\bigskip

    The case $\Delta=-\ft83$ is nothing but the well-known
AdS$_5\times S^5$ solution of type IIB supergravity, with no scalar
field.  The scalar potential for the case $\Delta=-\ft{12}{5}$ was
obtained from the $S^4$ reduction of the type IIA theory with a
non-vanishing 4-form field strength \cite{cllp,clpst2}, followed by an
$S^1$ reduction.  The potential for the first of the $\Delta=-2$ cases
was obtained from the $S^3$ reduction of ten-dimensional supergravity
with a non-vanishing 3-form field strength \cite{cllp,chamsab,clpst2},
followed by a reduction on $T^2$.  The scalar potential corresponding
to the second $\Delta=-2$ case arises from a new source.  It is
interesting that the D5-brane reduced on $T^2\times S^3$ and the
M5/M5-brane intersection reduced on $T^4\times S^2$ (or K3$\times
S^2$) give rise to the same scalar potential.  This may suggest some
duality relation between the quantum field theory living on the world
volume the D5-brane wrapped on $T^2$ and that of the M5/M5-brane
wrapped on $T^4$ or K3.

     In the previous subsection, we saw that the domain walls that
can trap gravity are those with $N=1$, $n\ge 3$ and $N=2$, $n=2$.
All the M-branes in $D=11$ and D$p$-branes in $D=10$ have $N=1$.
Clearly M-branes lead to domain walls in $D=4$ and $D=7$ that can trap
gravity since the internal sphere dimension $n$ is greater than 3.
For D$p$-branes, the corresponding lower-dimensional domain wall can
trap gravity only for $p\le 5$.  In section 4, we shall show this may
be related to the fact that a natural decoupling limit exists only
for D$p$-branes with $p\le 5$, but not for $p\ge6$.

\subsection{Two-scalar domain-wall solutions}

    In this section we shall consider the domain-wall solutions for
the two-scalar potentials obtained in section 3.2, with the second
scalar $\varphi$ no longer set to zero.\footnote{In the case of $b=0$,
the scalar potential would comprise only the massive breathing mode.
The associated domain-wall solution using this breathing mode was
obtained in \cite{bdlps}.  In \cite{clpbreath1,clpbreath2} it was used
in order to obtain a gravity-trapping model analogous to
Randall-Sundrum II (with one 3-brane), exploiting the fact that the
breathing mode is massive, and so it has a scalar potential with a
minimum, rather than a maximum. (The inclusion of a delta-function
source is still necessary.)  The use of the breathing mode was then
explored in the context of the Randall-Sundrum I scenario
(with two 3-branes \cite{rs1}) in \cite{alwis1,alwis2,dls}.}  We saw
in section 3.3 that the single-scalar solutions where $\varphi=0$ had
the interpretation, after lifting back to $d+n$ dimensions on $S^n$,
of being the near-horizon limits of isotropic $(d-2)$-branes.  In fact
we could have reversed the process, and derived the lower-dimensional
domain walls by reducing these near-horizon limits on $S^n$.  We shall
use the analogous inverse procedure now in order to obtain the
lower-dimensional two-scalar domain-wall solutions.

   To do this, we begin by considering the $(D-n-2)$-brane solution
(\ref{genbranes}) in $D$ dimensions.  This time, however, we shall
retain the constant ``1'' in the harmonic function $\hat H$.  Reducing
the metric on $S^n$, following the Ansatz (\ref{spherered}), we obtain
the $d$-dimensional solution
\bea
ds_d^2 &=& \hat H^{\fft{N}{d-2}}\, r^{\fft{2n}{d-2}}\, \eta_{\mu\nu}\,
dx^\mu\, dx^\nu + \hat H^{\fft{N\, (d-1)}{d-2}}\, r^{\fft{2n}{d-2}}\,
dr^2\,,\nn\\
e^{-\fft{2}{N\, a}\, \phi_1} &=& \hat H\,,\qquad e^{\fft{2(d-2)\, \a}{n}\,
\phi_2} = \hat H^{\fft{(d-1)\, N}{D-2}}\, r^2\,.
\eea
The metric can be recast into a conformal frame 
\be
ds_d^2 =  \hat H^{\fft{N}{d-2}}\, r^{\fft{2n}{d-2}}\, (\eta_{\mu\nu}\,
dx^\mu\, dx^\nu +dy^2)\,,
\ee
by introducing a new coordinate $y$, related to $r$ by $dy= \hat
H^{N/2}\, dr$, which implies
\be
y = r\, _2F_1\Big[\fft{1}{1-n}, -\ft12 N; 1+\fft{1}{1-n};
-\fft{Q}{r^{n-1}}\Big]\,.
\ee

   As $r$ tends to infinity, $y$ becomes proportional to $r$ and
therefore tends to infinity.  As $r$ approaches the horizon at $r=0$,
the coordinate $y$ tends to $-\infty$.  Thus we can introduce an
appropriate cut-off by changing coordinate from $y$ to $z$, defined by
\be
y= -c\, (1+k\, |z|)\,,
\ee
where $c$ and $k$ are positive constants.  By doing this we have
introduced a domain-wall, with a standard delta-function curvature
singularity, at $z=0$, which corresponds to some value of $r$ outside
the $(D-n-2)$-brane's horizon at $r=0$.  As $z$ tends to $\pm\infty$
the coordinate $r$ tends to zero, and so the horizons on either side
of the singular wall are mapped into the horizon of the
$(D-n-1)$-brane.  This is the desired two-scalar solution of the
Lagrangian that we obtained in section 3.2.  Clearly, it approaches
the same form as the previous 1-scalar solution as $z$ goes to
$\pm\infty$, since in these regions $r$ becomes small enough that the
``1'' in the harmonic function $\hat H$ becomes negligible.  It
follows that the characteristic structures of the associated
Schr\"odinger potentials for the two-scalar solutions will be
essentially the same as for the corresponding 1-scalar solutions
discussed in section 2.

\section{Domain-wall/QFT correspondence}

\subsection{Decoupling limit and the localisation of gravity}

    Although only the D3-brane has a near-horizon limit of
AdS$_5\times S^5$, which leads to the conjecture of the
AdS$_5$/CFT$_4$ correspondence, one expects that the world volumes of
other D-branes in ten dimensions should also describe certain quantum
field theories.  It was observed in \cite{towske} that for any
D$p$-brane in ten dimensions there exists a dual frame where the
metric becomes a direct product of AdS$_{10-n} \times S^n$, for $n\ne
3$, or M$_7\times S^3$, where M$_7$ is seven-dimensional Minkowski
space-time.  Since dimensional reduction of a D$p$-brane on $S^{8-p}$
gives rise to a domain wall in general, this leads to the conjecture
of a Domain-wall/QFT correspondence \cite{towske}.  In \cite{clpuniv},
ellipsoidal distributions of D$p$-branes were obtained, and it was
shown these solutions can be consistently reduced on the internal
transverse spheres to give rise to multi-scalar domain walls,
generalising the results of the Coulomb branch of the AdS/CFT
correspondence \cite{KLT,FGPW,BS,BSI,cglp,BBS}.  Here we shall
investigate the Domain-wall/QFT correspondence in the context of the
relation between the gravity-decoupling limit of a D-brane, the
trapping of gravity on the associated lower-dimensional domain wall.

         As we discussed in the previous sections, the Lagrangian
(\ref{genlag}) admits a $(D-n-2)$-brane solution, given in
(\ref{genbranes}).  There exists a dual frame $ds_{\rm dual}^2
=e^{-a\phi/(n-1)} ds_{\rm Einst}^2$, in which the near-horizon region
of the dual frame metric becomes
\be
ds_{\rm dual}^2 \sim r^{(n-1)\, N-2} \, dx^\mu\, dx_\mu + \fft{dr^2}{r^2}
+ d\Omega_{n}^2\,.
\ee
Thus when $(n-1)N=2$, corresponding to $\Delta=-2$, the metric is
M$_d\times S^n$; otherwise it is AdS$_d\times S^n$.

        In order to make sense of the Domain-wall/QFT correspondence,
it is necessary to examine whether there exists a natural decoupling
limit in which the gravitational constant can be set to zero, so that
the higher-dimensional $p$-brane limits to the domain wall.  To do so,
one needs to make a coordinate rescaling
\be
r = u\, (\ell_p)^\gamma\,,
\ee
and then send the Planck length $\ell_p$ to zero while keeping $u$
fixed.  In this limit, the constant ``1'' in the harmonic function
$\hat H$ can be dropped if $\gamma >1$, since the charge $Q$ has to
scale as $\ell_p^{n-1}$.  The natural decoupling limit exists if the
metric can now be expressed as $ds^2_{\rm dual} = \ell_p^2 \, d\td
s^2_{\rm dual}$ where $d\td s^2$ is independent of $\ell_p$.  For the
solutions given by (\ref{genbranes}), we find that this can be
achieved only when the following condition is satisfied:
\be
(n-1)\, N = \fft{2\gamma}{\gamma-1}\,.
\ee
Thus for $N=1$, which applies for all the D-branes in ten dimensions,
the requirement that $\gamma>1$ implies $n\ge 3$. (The $n=3$
case requires $\gamma=\infty$.)  In other words, the decoupling limit
exists naturally only for D$p$-branes with $p\le 5$, which have
$n\ge3$.  This condition for the existence of a decoupling limit
coincides precisely with the requirement that gravity be localised on
the brane as discussed in section 2.

   This coincidence may not be surprising.  It was shown that the
Randall Sundrum II wall with the AdS$_5$ bulk geometry can be
interpreted within an AdS$_5$/CFT$_4$ correspondence as a singular
domain wall source that cuts off the boundary of AdS$_5$
\cite{hver,witmaldunpub,gub1,gidkatran,noz,hawherrea,duffliu}. The
delta-function source in turn provides an ultra-violet cut-off in the
dual CFT$_4$.  In the case of dilatonic domain walls, discussed in
section 2, it should likewise be possible to view the domain walls
source for the gravity trapping solutions as an ultra-violet cut-off
for the dual QFT within the Domain-wall/QFT correspondence.  Thus the
criterion to localise gravity could have been expected to coincide
with the criterion for the decoupling of gravity in the complementary
picture of Domain-wall/QFT correspondence.

        The absence of a natural decoupling limit of D$p$-branes with
$p\ge 6$ is also related to the difficulties arising in M(atrix)
theory on a torus $T^n$ with $n\ge 6$ \cite{sen,seiberg}.  We showed
in this paper that the corresponding domain walls do not have
localised gravity, unlike the cases with $p\le 5$.  The difficulty may
also be related to the fact that the U-duality group becomes
exceptional for $D\le 5$, which requires the dualisation of
$(D-1)$-forms to enlarge the scalar coset \cite{cjlp}.

\subsection{One-loop corrections from the QFT}

    In the context of AdS/CFT, an interesting observation was recently
made \cite{duffliu}, to the effect that if one treats the
Randall-Sundrum II picture as a UV cut-off applied to the AdS/CFT
system, then the leading-order corrections to localised gravity on the
brane are in exact agreement with the corrections to the graviton
propagator induced by one-loop contributions from the super Yang-Mills
fields on the boundary.  One might expect, therefore, that it should
be possible to see an analogous complementarity in the conjectured
Domain-wall/QFT equivalence.  The technology for performing such a
detailed comparison is not yet sufficiently developed.  However, we
can determine what the general structure of the one-loop QFT
corrections would have to be if the conjectured correspondence were to
hold.

    In section 2.3 we determined the leading corrections to Newtonian
gravity at large distances $r$.  It was observed in \cite{duffliu}
that the $1/r^3$ correction in the Randall-Sundrum II case
($\Delta=-\ft83$) was exactly reproduced by the corrections to the
graviton propagator coming from the effect of closed super Yang-Mills
loops, which, in momentum space, takes the general form
\be
\Pi(p) \sim a\, \log\fft{p^2}{\mu^2} + b\,.
\ee
For the case $\Delta=-\ft{12}{5}$, we would instead need a propagator
correction of the form
\be
\Pi(p) \sim a\, p^2\, \log\fft{p^2}{\mu^2} + \cdots\,,
\ee
in order to yield, after a Fourier transformation, the required
$1/r^5$ leading-order correction to the Newtonian potential.
    
    For the $\Delta=-2$ examples, we saw in section 2.3 that the
leading-order correction to Newtonian gravity is of the Yukawa-like
form $e^{-k\, r/2}\, r^{-5/2}$.  We find that in this case the needed
corrections to the graviton propagator coming from one-loop effects in
the boundary QFT would be of the form
\be
\Pi(p) \sim \fft1{\Big((p^2+\ft14 k^2)^{\fft12} + \ft12 k\Big)^{\ft12}}\,.
\ee

\section{Conclusions}

   The primary goal of this paper was to study the localisation of
gravity for a more general class of domain walls, and to provide a
complementary description of these phenomena within the
Domain-Wall/QFT correspondence.

   For this purpose we studied thin, flat (BPS) domain walls in $d$
dimensions for which the asymptotic geometry is a vacuum with a
running dilaton. (The examples with asymptotic AdS space-times are
special cases within this class of solutions.)  The domain walls arise
as solutions of a $d$-dimensional bulk Lagrangian with a scalar field
whose potential is a pure single exponential $e^{b\, \phi}$, and where
a singular domain-wall source is introduced.  In particular, we
reviewed the space-time structures of these solutions with the focus
on those where the domain-wall source has positive tension with no
naked singularities, and where the bulk Lagrangian determining the
space-time on the two sides of the wall is the same.  Thus the
domain-wall geometry is $Z_2$ symmetric.  (The domain wall with
asymptotic AdS \cite{rs2} is a particular example within this class of
solutions.)  In order to have a trapping of gravity on the wall, the
tension must be positive, and in addition the coupling constant $b$ in
the scalar exponential potential must lie within certain bounds.  

    The exponential potentials responsible for such gravity-trapping
domain-wall solutions can arise from certain sphere reductions in
M-theory or string theory, whilst a generalised Scherk-Schwarz
reduction on Ricci-flat internal space, which also gives a pure
exponential potential, does not yield a value for $b$ in the range
necessary for the trapping of gravity.  Specifically, we classified
such examples of sphere compactifications and focused on the explicit
examples that yield effective theories in $d=5$.

    The gravity-trapping domain-wall solutions can be lifted back on
these internal spheres to $D=11$ or $D=10$.  Their bulk geometries
turn out to be the near-horizon regions of M-branes or D$p$-branes
(with $p\le 5$), with the domain wall itself located at some distance
outside the horizon, and the two horizons of the domain-wall solution
corresponding to the horizon of the M-brane or D$p$-brane.  It is
intriguing that these are precisely the near-horizon geometries of the
branes for which a natural gravity-decoupling limit exists, and so
these are the examples for which a Domain-wall/QFT correspondence can
be established.  Conversely, the D$p$-branes with $p\ge6$, which do
not have a natural decoupling limit, are associated with lower-domain
walls that cannot trap gravity.  This provides strong evidence to
suggest that the localisation of gravity on the domain wall and the
existence of a Domain-wall/QFT correspondence are closely related.

      Note if we remove the delta-function source for the
gravity-trapping domain walls, then the metric from one side runs from
the null singularity at $-\infty$ to a (non-singular) boundary at some
finite $z$.  Thus the introduction of the singular delta-function
source, before the boundary of this space-time is reached, can now be
viewed within the Domain-wall/QFT correspondence as the introduction
of an ultra-violet cut-off in the boundary field theory, thus
generalising the complementarity between the Randall-Sundrum II
scenario and the AdS/CFT correspondence
\cite{hver,witmaldunpub,gub1,gidkatran,noz,hawherrea,duffliu}.

     An important test of this complementarity is to see if the
leading-order corrections to Newtonian gravity on the dilatonic
domain-walls are in agreement with the predictions from the
corrections to the graviton propagator induced by one-loop QFT
effects, which again generalises the studies within the AdS/CFT
correspondence.  Within this framework we analysed the linearised
equation describing gravity fluctuations, and obtained the spectrum.
In one example, where the spectrum has a mass gap, we obtained the
exact analytic form for the associated Green function.  In all the
examples where gravity is trapped on the domain wall we obtained the
leading-order corrections to Newtonian gravity, and from these we
deduced the necessary forms of the one-loop boundary-QFT contributions
to the graviton propagator.  It would be interesting to study this in
more detail, by comparing with detailed one-loop results from the
boundary QFT, to obtain further tests of the proposed complementarity.

\section*{Acknowledgements}

    We are grateful to Jianxin Lu for useful discussions on the
gravity-decoupling limit of D$p$-branes, and to Konstadinos Sfetsos
for discussions on normalisation conditions for fluctuation modes in
gravitational backgrounds.  H.L. and C.N.P. are grateful to CERN for
hospitality during the course of this work.

\appendix

\section{Exact Green function for $\Delta=-2$ gravity perturbations}

   The Green function satisfies the equation
\be
-\fft{d^2G(x,z;x',z')}{dz^2} -\square_4\, G(x,z;x',z') + 2U\,
G(x,z;x',z') =\delta^5(x,z; x',z')\,,
\ee
where for $\Delta=-2$ the Schr\"odinger potential $U$ is given by the
second line in (\ref{upot}).  We saw earlier that the massless bound
state wave-function $u$ and the massive continuum wavefunctions $u_q$
are given by
\be 
u(z)=e^{-\fft12 k\, |z|}\,,\qquad u_q(z)= k\, \sin q|z| - 2q\, \cos q|z|
\,,
\ee
where $u_q$ corresponds to mass $m= \sqrt{q^2 + \ft14 k^2}$.  The
retarded Green function is then given by
\bea
G(x,z;x',z')_R &=& \int\fft{d^4 p}{(2\pi)^4}\, e^{\im\, p\cdot (x-x')}
\, \Big\{ \fft{k\, u(z)\, u(z')}{2(\vec p^2 - (\omega-\im\, \ep)^2} 
\nn\\
&&+ \int_0^\infty dq\, \fft{u_q(z)\, u_q(z')}{\pi\, (4q^2+ k^2)\, [\vec
p^2 - (\omega-\im\, \ep)^2 + q^2 + \ft14 k^2]}\Big\}\,.
\eea
which can be evaluated explicitly.  The static Green function is given
by integrating over time, yielding $G(\vec x, z; \vec x', z')=
\int_{-\infty}^\infty  dt'\, G(x,z;x',z')$.

   Taking $z=z'=0$, we eventually obtain the following expression for
the static Green function $G(\vec x; \vec x') =G(\vec x, 0;\vec x',
0)$:
\bea
G(\vec x; \vec x') &=& \fft{k}{8\pi\, R} + 
\fft{e^{-\ft12 k\, R}}{4\pi^2\, R}\, \int_0^\infty dy 
\fft{(y(y+k))^{1/2}}{y+\ft12 k}\, e^{-y\, R}\,,\nn\\
&=& \fft{k}{8\pi\, R} + 
\fft{k}{8\pi^2\, R}\, K_1(\ft12 k\, R)\nn\\
&& +
\fft{k^2}{32\pi} \, \Big( K_0(\ft12k\, R)\, L_{-1}(\ft12 k\, R) 
+K_1(\ft12k\, R)\, L_{0}(\ft12 k\, R) \Big) -\fft{k}{16\pi\, R}\,,
\label{green}
\eea
where $R=|\vec x-\vec x'|$, and $L_{n}(x)$ is the modified Struve
function.  This is a special case of the generalised hypergeometric
function $_pF_q$ for $p=1$, $q=2$:
\be
L_{-1}(x) = \fft{2}{\pi}\,  _1F_2[1; \ft12,\ft32; \ft14 x^2]\,,\qquad
L_{0}(x) = \fft{2x}{\pi}\,  _1F_2[1; \ft32,\ft32; \ft14 x^2]\,.
\ee
Note that the first term in (\ref{green}) is the usual one, which
comes from the massless mode $u(z)$; all the remaining terms come from
the massive modes $u_q(z)$.  The form of this result, and in
particular the integral associated with the massive modes, is
identical to that obtained in section 2.3. Although the final term in
(\ref{green}) is of the same form as the leading-order result from
$u(z)$, it is appropriate to keep it distinct since it is really just
cancelling an equal and opposite term that resides in the products of
Bessel and Struve functions in a large-$R$ expansion.

    At large $R$, we therefore have
\be
G(\vec x; \vec x') = \fft{k}{8\pi\, R} + \fft{k^2\, e^{-\ft12k\, R}}{4
\pi^{3/2}}\, \Big[ \fft1{(k\, R)^{5/2}} - \fft{9}{4(k\, R)^{7/2}} +
\fft{345}{32 (k\, R)^{9/2}} +\cdots\Big]\,.
\ee
This is in precise agreement with the correction to the Newtonian
potential discussed in section 2.3.

    At small $R$, we find
\be
G(\vec x; \vec x') = \fft1{4\pi^2\, R^2} + \fft{k}{16\pi\, R} -
\fft{k^2}{32\pi^2}\, \log(\ft12 k\, R) +\cdots\,.
\ee
Thus, as expected, the gravity becomes effectively five-dimensional at
small length-scales.

\end{document}